\documentclass[conference]{IEEEtran}

\usepackage{graphicx}
\graphicspath{{images/}}
\usepackage{layout}

\usepackage[minbibnames=1,style=ieee, backend=biber, sorting=none]{biblatex}
\newcommand{\XPTO}{the proposed system}
\begin{document}

\title{CapillaryX: A Software Design Pattern for Analyzing Medical Images in Real-time using Deep Learning}

\author{\IEEEauthorblockN{Maged Abdalla Helmy Abdou, Paulo Ferreira and Eric Jul}
\IEEEauthorblockA{Department of Informatics\\
University of Oslo, Norway}
\and
\IEEEauthorblockN{Tuyen Trung Truong}
\IEEEauthorblockA{Department of Mathematics\\
University of Oslo, Norway}}

\maketitle
\thispagestyle{plain}
\pagestyle{plain}

\begin{abstract}
Recent advances in digital imaging, e.g., increased number of pixels captured, have meant that the volume of data to be processed and analyzed from these images has also increased. Deep learning algorithms are state-of-the-art for analyzing such images, given their high accuracy when trained with a large data volume of data. Nevertheless, such analysis requires considerable computational power, making such algorithms time- and resource-demanding. Such high demands can be met by using third-party cloud service providers. However, analyzing medical images using such services raises several legal and privacy challenges and do not necessarily provide real-time results. This paper provides a computing architecture that locally and in parallel can analyze medical images in real-time using deep learning thus avoiding the legal and privacy challenges stemming from uploading data to a third-party cloud provider. To make local image processing efficient on modern multi-core processors, we utilize parallel execution to offset the resource-intensive demands of deep neural networks. We focus on a specific medical-industrial case study, namely the quantifying of blood vessels in microcirculation images for which we have developed a working system. It is currently used in an industrial, clinical research setting as part of an e-health application. Our results show that our system is approximately 78\% faster than its serial system counterpart and 12\% faster than a master-slave parallel system architecture.

\end{abstract}

\IEEEpeerreviewmaketitle

\section{Introduction} 
\label{intro}

Early attempts to address the problem of running demanding 
computational algorithms in tightly constrained environments 
emerged in the 1980s~\cite{duncan1990survey}. 
Performance limitations became apparent with the rise of processing 
big data using deep learning (DL) techniques
because DL requires large amounts of computational
power~\cite{zhang2018survey,thompson2020computational}. 
Such limitations included the under-utilization of the available 
computing resources to execute processes introducing undesirable 
delays~\cite{pumma2017parallel}. 
These limitations are still prominent when real-time results are desired in tightly constrained environments (i.e., clinical environments).
Furthermore, using third-party cloud services to rent computing resources is risky due to General Data Protection Regulation (GDPR)~\cite{rumbold2017effect}. These regulations effectively limit clinicians to local computing resources, such as laptops and PCs approved for use at hospitals.

This paper aims to design, implement, and evaluate a software package that can analyze medical images using deep learning in a local environment as to mitigate the risk of breaching GDPR rules while still getting results in real-time.
We focus on a specific industrial, medical case study: the quantification of blood vessels in microcirculation images captured by using in-clinic, hand-held cameras with microscope lenses.
The quantified value is called capillary density or blood vessel density.
This value is of high clinical relevance because the fluctuation of this value can be used as an early marker to indicate an organ failure, and the severity of the change might predict the chances of the patient surviving~\cite{JS_Parker2019-vp, Nama2011-cz, fagrell1997microcirculation, Houben2017-bz, Wester2014-tz,Lopez2015-se,hernandez2013microcirculation}.

The requirements of our system were established by interviewing 
a set of medical doctors and surgeons who spent several years in the microcirculation analysis field (associated with ODI Medical AS, a MedTech company responsible for the e-health industrial application). 
The main requirements for a production-grade system for the quantification of blood vessels analysis captured by a real-time camera are:

\begin{enumerate}
\item the system must be able to analyze a microcirculation image (1920x1080) in real-time (one second or less);
\item the system must have low power consumption so that it can be used in battery-powered devices in hospitals; and
\item the system must be built on top of a popular, widely used programming language and framework (e.g., Python and Tensorflow) running on standard hardware.
\end{enumerate}

To the best of our knowledge, no previous work on microcirculation analysis reported using parallel frameworks to calculate the capillary density in under 1 second for a frame with a resolution of 1920x1080 on a CPU using deep learning with an accuracy of $\sim$85\%. The medical doctors proposed this accuracy value to outperform the accuracy achievable by a trained clinician.
Previous systems in the literature that achieve a comparable or higher accuracy needed a GPU that is not available in typical low-power computers approved for use in hospitals.
The developed system runs in an industrial, clinical environment on a standard low cost computer utilizing all the available resources and meets the requirements listed above.
This paper does not focus on developing the deep learning algorithm but rather the deployment of the deep learning algorithm. The algorithm used in this paper achieves an accuracy of $\sim$85\%, and is described in a previous
paper~\cite{helmy2021capillarynet}.

In Section~\ref{related_work}, we present the work related to our paper including a literature survey on the relevant parallel frameworks and existing systems that were built to analyze microcirculation images.
In Section~\ref{proposed_system}, we present the proposed architecture for our package along with two baseline systems that we used to benchmark our proposed architecture against.
In Section~\ref{implementation}, we present how we implemented our system.
In Section~\ref{evalaution}, we present the evaluation criteria that have been used to evaluate our system and benchmark our proposed system against a baseline serial system and a baseline parallel system with the presented criteria and discuss our results.
In Section~\ref{conclusion}, we present our conclusion.

\section{Related Work}
\label{related_work}

This section presents the literature review on current parallel frameworks and existing systems built to calculate capillary density.

\subsection{Parallel Frameworks}

Hadoop~\cite{holmes2012hadoop} gained recognition in 2004 and provides a framework for distributed storage and the processing of big data. 
It splits large blocks of data into a Hadoop Distributed File System (HDFS) 
which is based on Google’s file system (GFS) and stores data across 
clusters~\cite{borthakur2008hdfs,shvachko2010hadoop}. 
HDFS uses data locality, allowing clusters and nodes to manipulate data, making it faster than conventional high-performance computing~\cite{dean2008mapreduce}. 
 
MapReduce then processes the data stored on HDFS~\cite{dean2010mapreduce}. 
MapReduce has a master job tracker and one per cluster to schedule jobs, manage resources, and re-execute processes when a node fails~\cite{dean2008mapreduce}. 
HDFS and MapReduce are two modules built to store and process big data reliably. 
However, the main drawback of Hadoop is that it cannot deal 
with big data real-time stream processing; therefore, Apache Spark was 
introduced~\cite{saouabi2017comparative} was introduced in 2010.

Some benchmarks show that Spark is three 
times faster than Hadoop ~\cite{gopalani2015comparing}. 
This increase is because Spark can load and process data using RAM instead of 
the two-stage access paradigm introduced by MapReduce~\cite{saouabi2017comparative}.
Spark outshines MapReduce when it comes to real-time 
processing~\cite{aziz2018real,van2007python}. 
Furthermore, the ease of programming on 
Spark with Scala~\cite{odersky2004overview}, Java ~\cite{arnold2005java} 
and Python~\cite{van2007python} makes it relatively easy to adapt 
instead of MapReduce, which can be programmed only in Java. 
Spark provides a unified processing system instead of several isolated 
applications that do not share the state amongst each other~\cite{spark2018apache}.
Although Spark was designed to outperform MapReduce processing, its 
the fundamental limitation is the complexity involving asynchronous execution 
and the compatibility issues introduced when integrating it into the deep 
learning lifecycle~\cite{moritz2018Ray}. 

Dask~\cite{rocklin2015dask} was introduced in 2014 and is a parallel
computing library that uses dynamic task scheduling to leverage 
multi-core processors and High-Performance Computing (HPC) clusters.
Instead of loading all data into RAM, Dask pulls data into RAM in 
chunks and throws away intermediate values as soon as possible, 
freeing more memory to process more data~\cite{rocklin2015dask}.
While Spark can be seen as an 
extension to the MapReduce paradigm, Dask is a generic task scheduling system
that handles complex dimensional arrays~\cite{dugre2019performance,mehta2016comparative,dask1}.
Both Dask and Spark leverage acyclic graphs, but the map stage of Dask can represent more complex algorithms than Spark~\cite{nishihara2017real}. 
Thus, Dask can parallelize sophisticated algorithms without excess memory usage~\cite{mehta2016comparative}.
Moreover, Spark does not natively support multi-dimensional arrays as Dask does~\cite{dask1,li2012performance}. 
This advantage makes Dask lightweight and smaller than Spark, and Dask integrates natively with the numeric Python ecosystem.
However, Dask is not fully compatible with TensorFlow, and deep learning algorithms as the framework focuses on Data science libraries like Pandas and Numpy. 


Orleans~\cite{bykov2011orleans} is an actor system that provides highly 
available concurrent distributed systems. The main drawback is how the 
system reacts to a data failure event. Developers must manually create 
checkpoint actor states and intermediate responses to restore 
stateful actors~\cite{bernstein2014orleans}. While this does not 
affect the performance, it can bring some overhead when developing a system 
to handle failure events.

Tensorflow~\cite{tensorflow1} is an ecosystem of machine learning and deep 
learning tools that leverage CPUs and GPUs while training. However, it 
provides limited support when deploying it to serve users because it does not 
fully support responses when a task is completed or when a fault is detected.
One way to perform this activity is to wrap the Tensorflow Model in a flask 
service and serve the model~\cite{aslam2015efficient}. 
However, this becomes unmanageable when scaling with different models.
Tensorflow serving~\cite{olston2017tensorflow} was introduced to deploy models in 
production environments but has to be used in conjunction with traditional 
web servers, which introduces additional latency. 

With the introduction of deep learning techniques~\cite{litjens2017survey}, 
which consists of several millions of parameters to compute, wrapping deep 
learning models with traditional servers is no longer sufficient.
Compared to traditional models, deep learning models are computationally intensive 
and have a response time of tens of a millisecond or greater~\cite{han2017efficient}; 
thus, there is a need for efficient parallelizing to reduce the response time. 

Frameworks such as MapReduce~\cite{dean2008mapreduce} and 
Spark~\cite{zaharia2012resilient} are not suitable for models serving 
in real-time because they were designed and built for batch processing.
Furthermore, they are not suitable for large numbers of small 
transactions because of the considerable time overhead that they require 
for instantiation. Dask~\cite{rocklin2015dask} and Tensorflow~\cite{tensorflow1} provide a complex 
and very little support for model serving~\cite{moritz2018Ray}. 

It is possible to set up different parts of different 
frameworks together to have a system that can serve a deep learning model. 
However, the compatibility and maintenance of these different frameworks increase the technical complexity. 
Unfortunately, deploying deep learning models into production is still not a straightforward endeavor.

\begin{figure*}[t]
\center
\includegraphics[width=\textwidth]{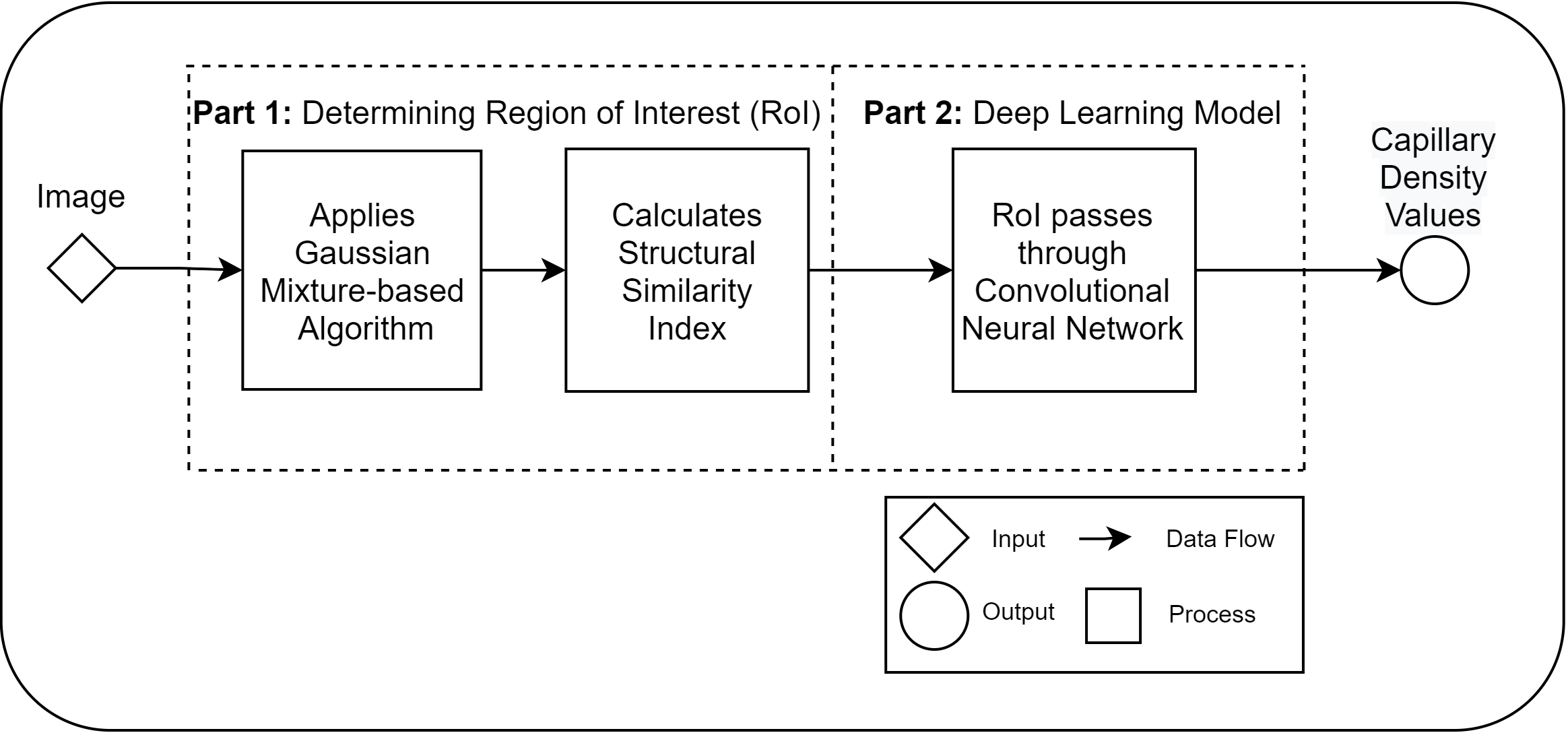}
\caption{The diagram shows the code encapsulated in a core on a computer.
This code is replicated across each core to achieve parallelism.
This code calculates the capillary density from a microcirculation image. 
The architecture consists of two parts: i) first determining the RoI using traditional computer vision algorithms and ii) then using deep learning to classify if the RoI contains a capillary}
\label{arch1}
\end{figure*}

\begin{figure*}
\centering
\begin{tabular}{cc}
\includegraphics[width=0.5\textwidth]{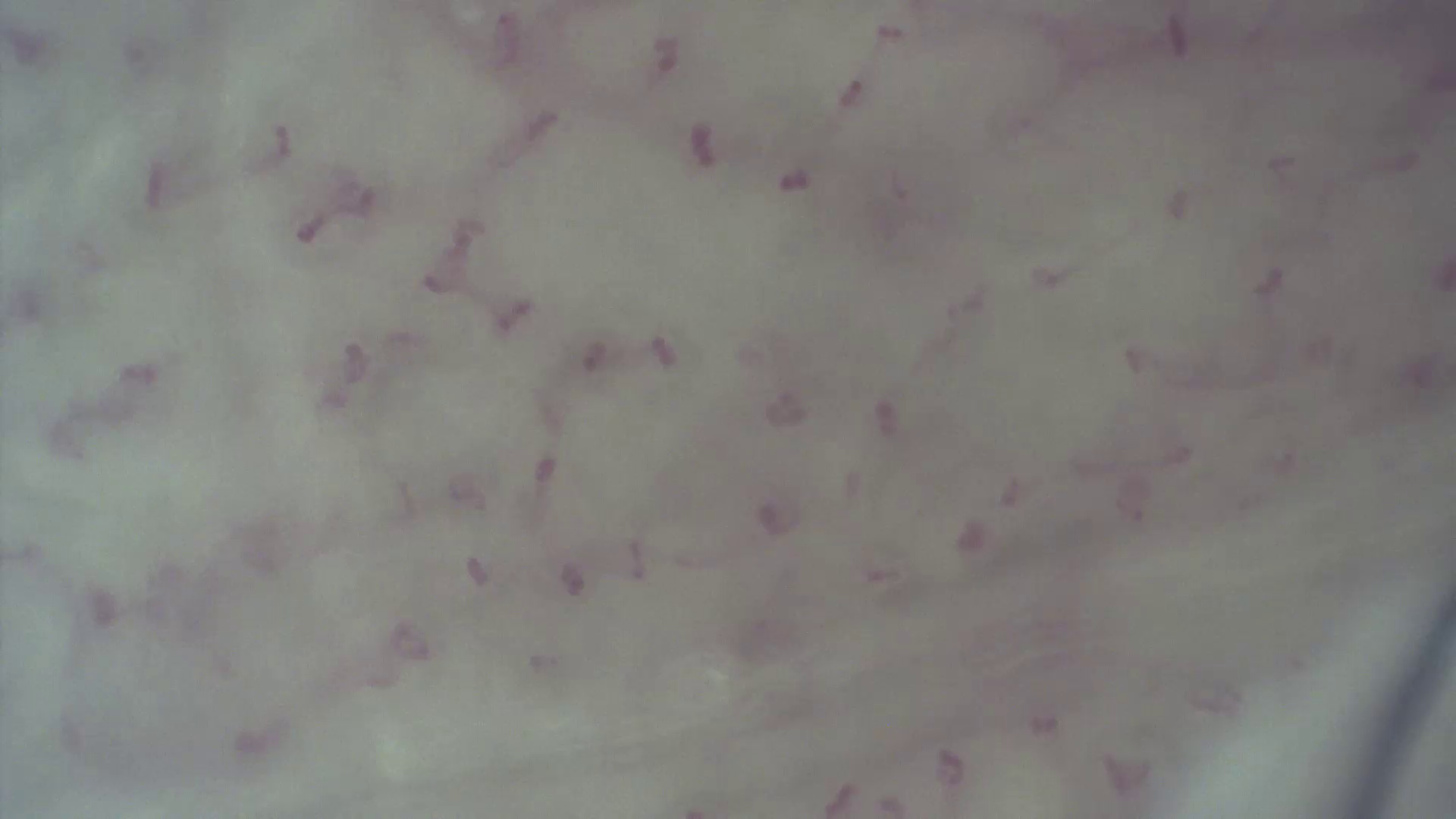}
\label{0_frame} 
& 
\includegraphics[width=0.5\textwidth]{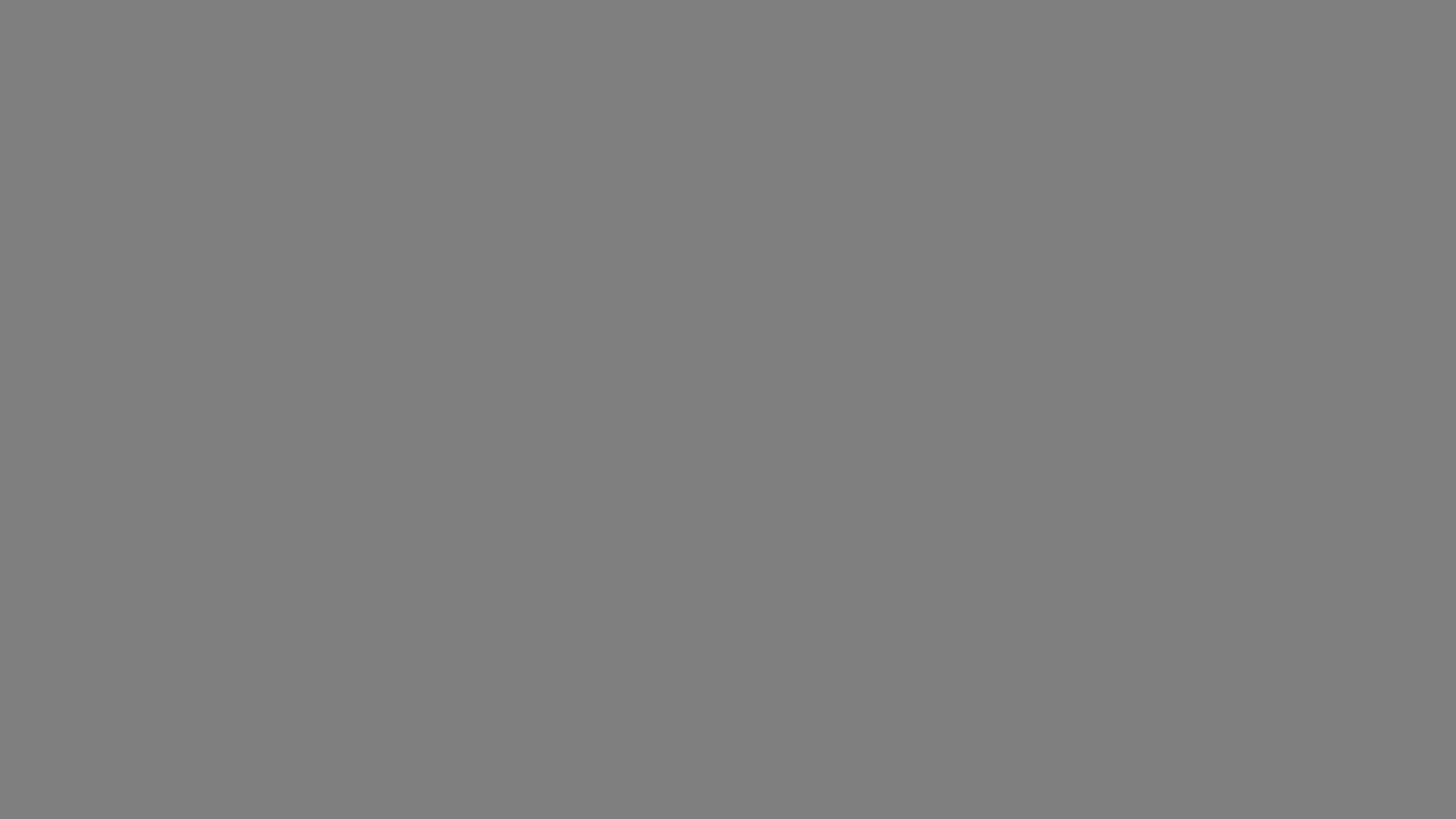} 
\label{1_image1_gray} \\
(a)     & (b)     \\
\includegraphics[width=0.5\textwidth]{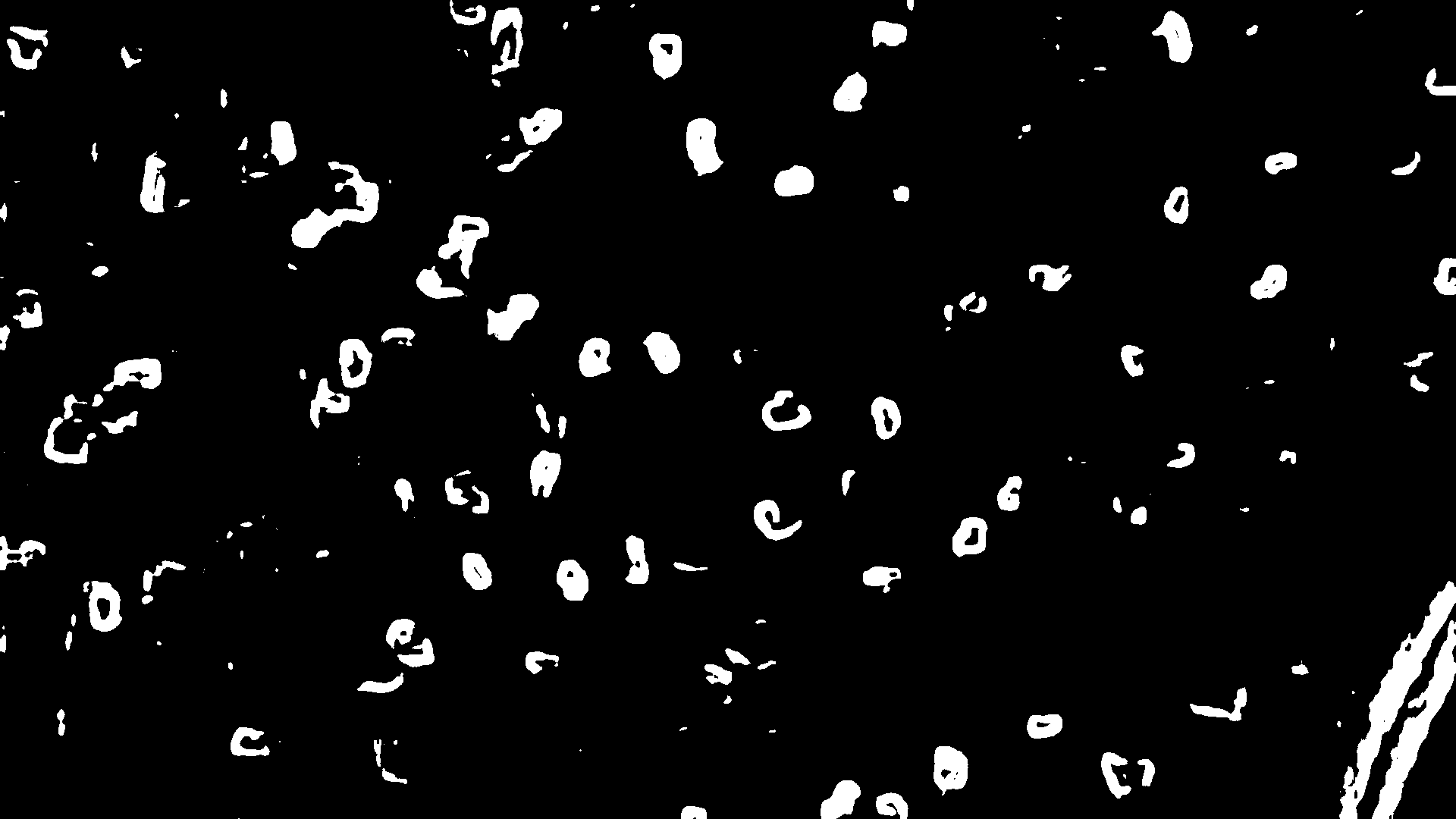}
\label{2_diff}  
& 
\includegraphics[width=0.5\textwidth]{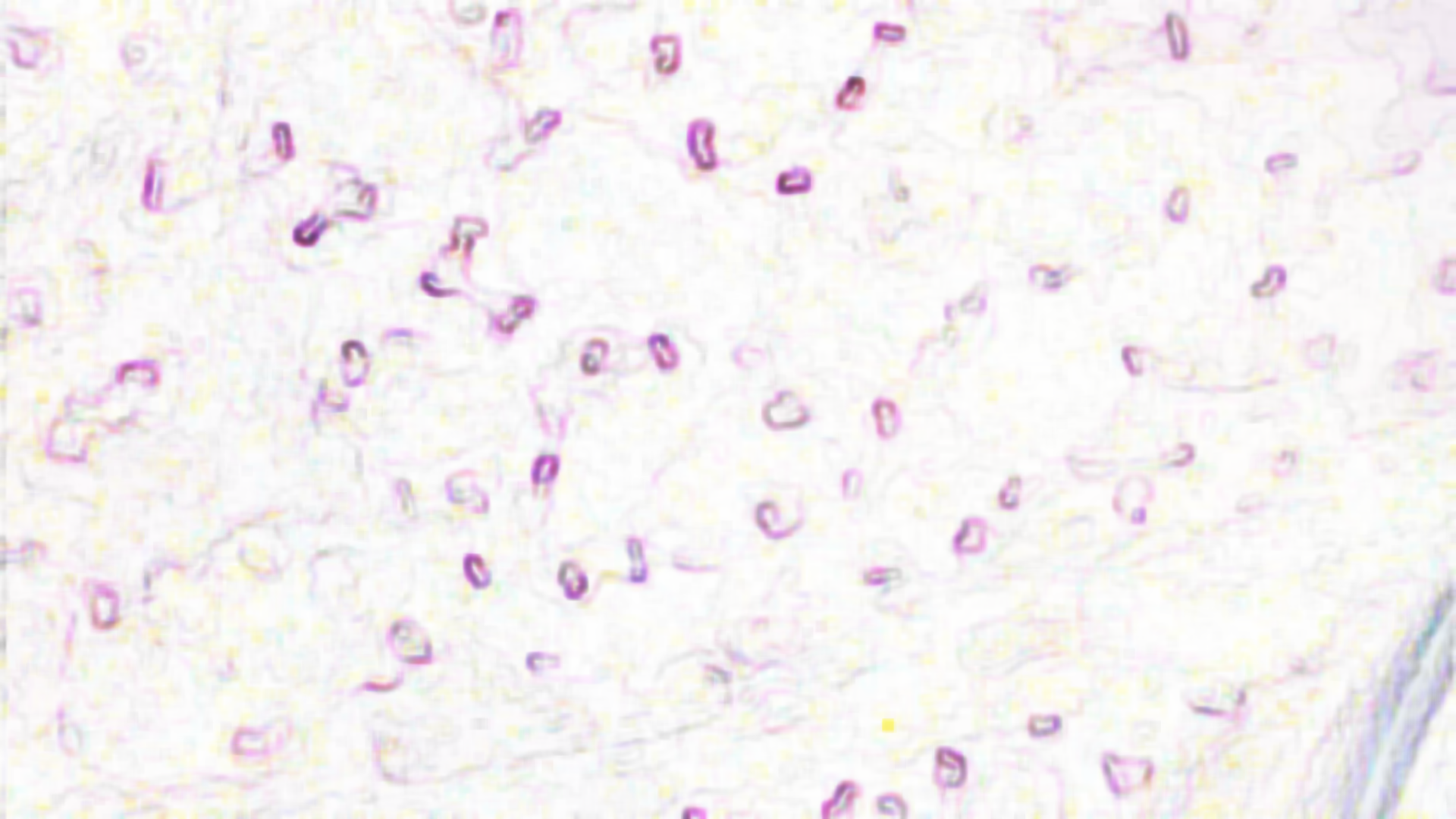}
\label{3_segmented_image} \\
(c)     & (d)     \\
\includegraphics[width=0.5\textwidth]{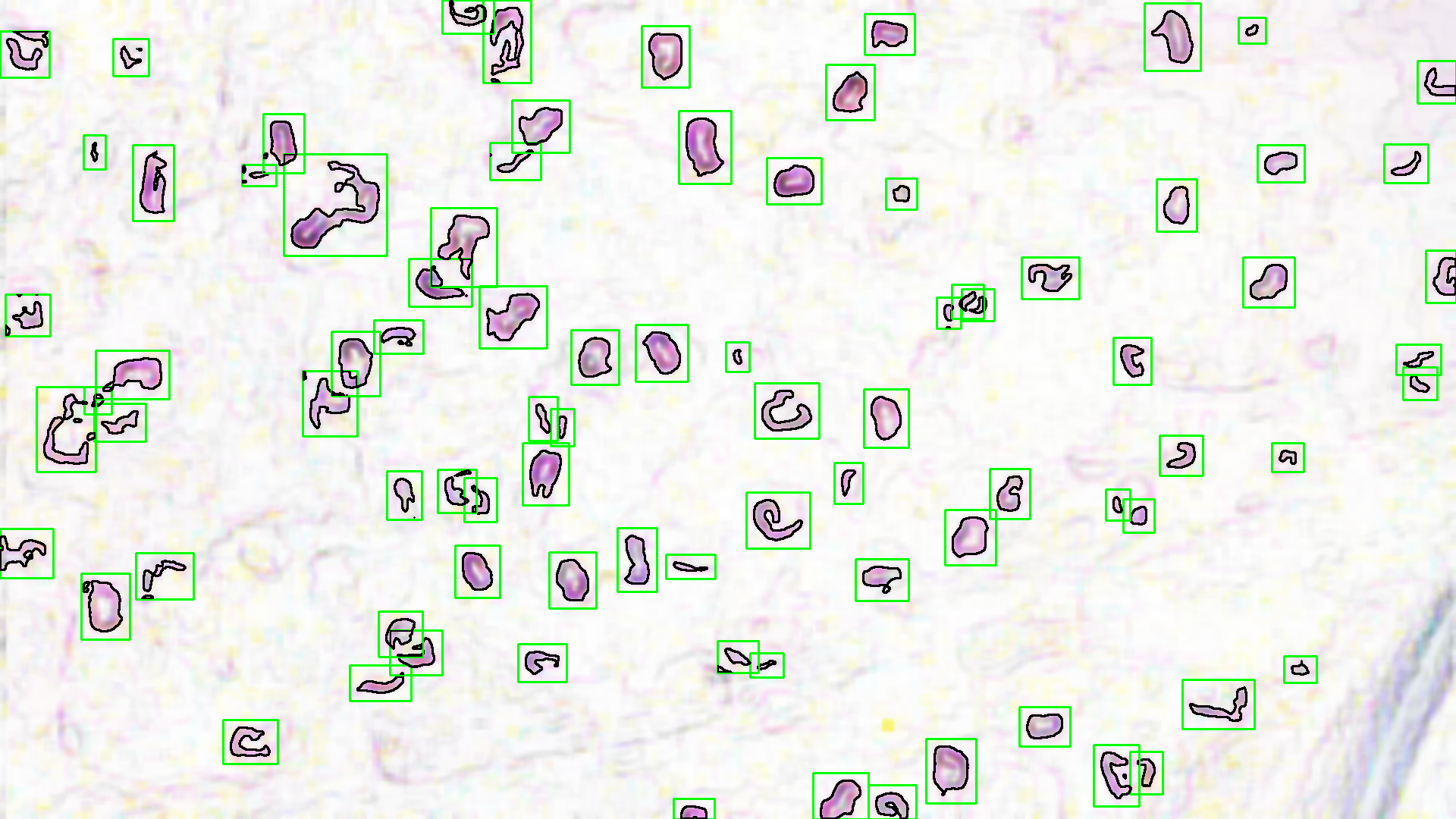}
\label{4_coloured_image} 
& 
\includegraphics[width=0.5\textwidth]{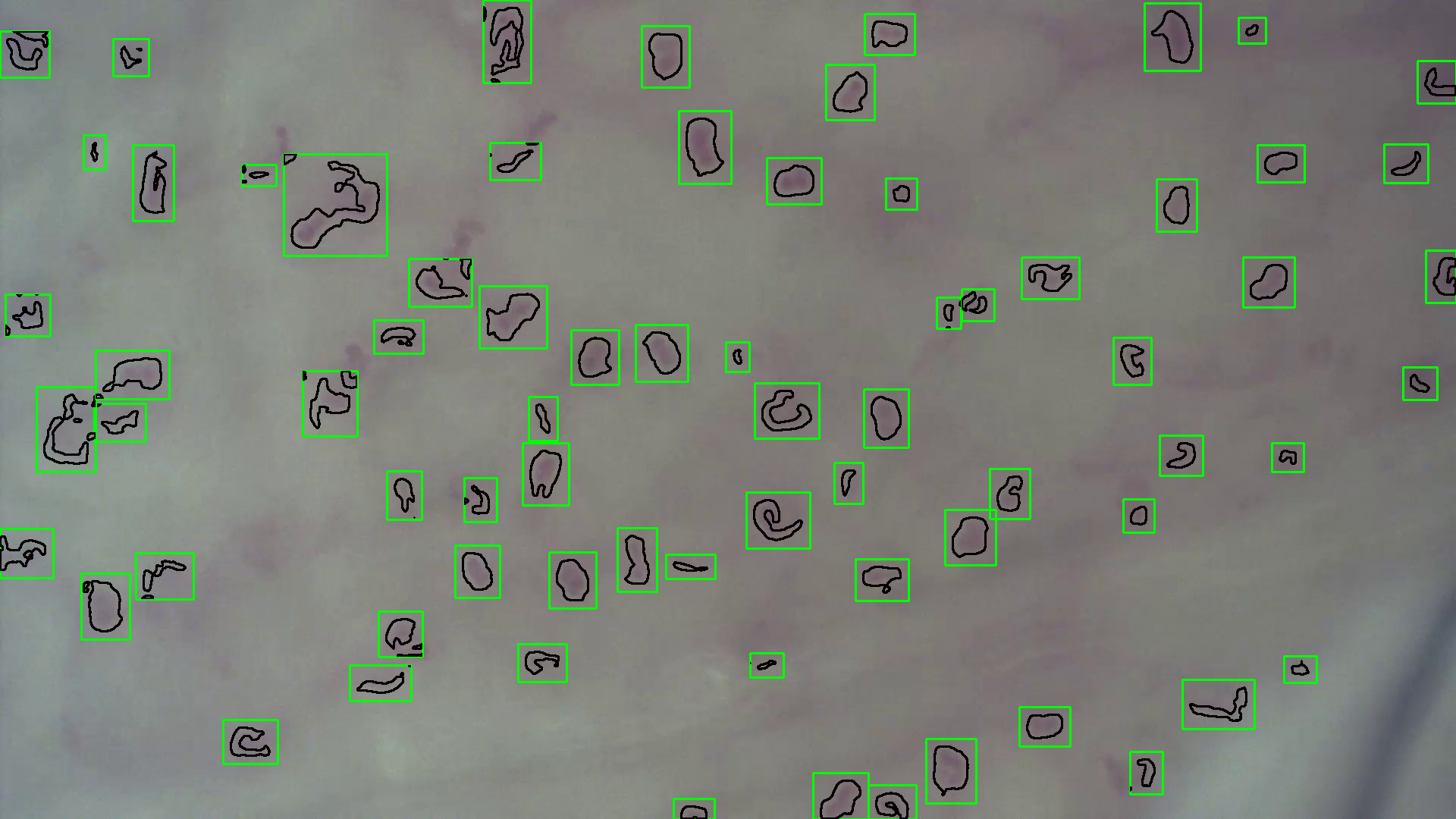}
\label{5_coloured_image} \\
(e)     & (f)     \\
\end{tabular}
\caption{
(a) This presents a sample of a microcirculation image that is taken as an input to the system 
(b) The background image calculated using a Gaussian Segmentation Algorithm 
(c) The segmented area formed by calculating the difference between the original image and the background
(d) The Structural Similarity Index calculated from the original frame and the background image 
(e) The modified image with the capillary area highlighted in black encapsulated within the green bounding box 
(f) The original image with the capillary area highlighted in black encapsulated within the a bounding box}
\label{example_image} 
\end{figure*}

\subsection{Existing Microcirculation Analysis Systems}

This section presents the current work on systems that calculate capillary density from microcirculation images.

As briefly mentioned at the end of the introduction, none of the existing works mentioned on microcirculation analysis reported using parallel frameworks to calculate the capillary density in under $\sim$1 second\ for a frame with a resolution of 1920x1080 on a CPU using deep learning with an accuracy of  $\sim$85\%. 
Those who exceeded this accuracy used a GPU which is not readily available in a clinical environment.

Cynthia Cheng et al.~\cite{cheng2015reproducible} takes a three-step approach to quantify capillary density.
First, they apply an image enhancement process to darken the capillaries and lighten the background.
They then flatten the image using 2D filters and raise the image's contrast.
The image is then despeckled using a 7x7 filter.
They then adjust the histogram of the image to a best-fit model
The second step involves manually selecting the capillary as a target object.
They then select the background as a reference.
The algorithm then selects the rest of the capillaries and excludes the images.
A macro is then created from this process, which can be applied to other images with similar characteristics.
As described, this involves several steps, including the manual user intervention; therefore cannot provide results in less than 1 second.

A. Tama et al.~\cite{tama2015nailfold} uses binarization followed by skeleton extraction and segmentation to quantify the capillaries. 
The first step involves extracting a reference image.
The image has to be then manually cropped by the user.
The green channel is then extracted from the image to have the highest probability of vessels in it.
They then apply a top-hat transform to remove unevenness in the background.
They then apply the Wiener filtering, a lowpass filter followed by Gaussian smoothing.
They then apply Otsu thresholding to segment the image from the background and apply a skeleton extraction method to quantify the capillary.
The authors do not report the speed needed to perform these steps.

Sherry G.Clendenon et al.~\cite{clendenon2019simple} uses a manual method to segment the microvascular structure. The authors do not report the speed or accuracy of their method.

Pavle Prentašic et al.~\cite{prentavsic2016segmentation} used a custom neural network to segment the foveal microvasculature. Their neural network consists of three Convolutional Neural Network (CNN) blocks coupled with max-pooling and a dropout layer followed by two dense layers. They reported accuracy of 82.4\% at 2 minutes.

R Nivedha et al.~\cite{nivedha2016classification} used a non-linear Support Vector Machine~\cite{noble2006support} to classify images.
They first started by extracting the green channel since it contains the relevant information to detect blood vessels.
They then performed manual cropping and used adaptive histogram equalization to improve the image's contrast.
They then used image enhancement to segment the image using a Gaussian filter followed by OTSU thresholding.
They then used PrincipalComponent Analysis(PCA) to extract the features.
A Support Vector Machine then performed the classification.
They reported accuracy of 83.3\% but not the time needed for automated analysis.

KV Suma et al.~\cite{suma2017fuzzy} used Fuzzy Logic Kernels to classify the images.
They started by Fuzzification of the input, followed by the Application of the Fuzzy operator, then aggregating the consequents across the rules, ending with results.
They reported an overall accuracy of 83.3\% but not the time needed for automated analysis.
In their next paper~\cite{suma2018novel}, they experimented with different types of machine learning techniques, including Random Forests Classifier, Multinomial Logistic Regression, and CNNs. However, they do not report the timing needed for classifying the blood vessels.

Perikumar Java et al.~\cite{javia2018machine} used a custom form of ResNet18~\cite{he2016deep} to quantify capillaries.
They used a 10-layer architecture and resized the images to input 224x224x3.
They applied the Adam optimizer and cross-entropy as loss metrics.
They trained the NVIDIA GeForce GTX Titan X algorithm and used the PyTorch library.
They reported accuracy of 88\%.
However, such an algorithm is not suitable for a clinical environment due to the high-end GPU required to run it.

F Ye et al.~\cite{ye2020vivo} utilized the concept of transfer learning and used the Inception Single Shot Multibox Detector (SSD) ~\cite{8026312} to build their neural network.
They build their system using Python and Tensorflow with an image resolution of 744 × 482 pixels.
They applied data augmentation to the image to increase the number of datasets.
The SSD architecture requires GPU to produce results in real-time, making it unsuitable to be used in a clinical requirement with only CPUs available.

YS Hariyani et al.~\cite{hariyani2020capnet} used U-net architecture combined with a dual attention module. 
They introduced a new method called DA-CapNet, which can analyze microcirculation images.
It consists of the encoder and decoder parts.
The encoder downsamples the dimension of the information in an image while increasing the number of channels. This step increases the spatial information dimension.
They then combine it with a dual attention module which increases the accuracy.
The dual attention uses the squeeze and excitation process to extract the blood vessels in the image.
The authors resized the image to 256×256 to reduce the processing time and used a Gaussian threshold method with a median blurring filter of kernel size five.
The authors reported accuracy of 64\% but not the time taken for analyses.

G Dai et al.~\cite{dai2020exploring} used a custom neural network similar to Pavle Prentašic et al. for segmentation. However, G Dai et al. used five CNN blocks instead of three. Hang-Chan Jo et al.~\cite{jo2021quantification} used a Attention-UNet architecture~\cite{oktay2018attention}.
Their method starts by using the CLAHE method and computes several histograms.
They then apply the Gamma correction and pass it to the deep neural network.
The reported accuracy was 73.20\%, but not the time is taken for analysis.

\section{Proposed System}
\label{proposed_system}
This section presents the system's architecture to analyze medical images in parallel, specifically, to calculate the capillary density in a microcirculation image. We start by presenting the DL part (which is based on OpenCV~\cite{bradski2000opencv} and Tensorflow~\cite{tensorflow1}) and the architecture of our system's parallel part (which is based on Ray~\cite{nishihara_moritz}).

\subsection{The Deep Learning Algorithm part of the Proposed System}

The outline of the deep learning architecture is shown in
Figure~\ref{arch1}. 
It consists of two main parts: 
i) determining the regions of interest (RoIs) where capillaries 
might exist, and 
ii) using a CNN for predicting whether these RoIs contain 
a capillary or not. 

The original frame is shown in Figure~\ref{example_image}a.
The position of the capillaries is determined by first removing the background from the original frame using a Gaussian Mixture-based Background/Foreground Segmentation Algorithm ~\cite{kaewtrakulpong2002improved}. 
The background removed is shown in Figure~\ref{example_image}b. 
The structural similarity index measure (SSIM) ~\cite{wang2009mean,wang2004image} is applied between the original frame 
shown in Figure~\ref{example_image}a 
and background image shown in Figure~\ref{example_image}b 
resulting in Figure~\ref{example_image}c and Figure~\ref{example_image}d. 
Bounding boxes are formed around the red areas using OpenCV contour method ~\cite{opencv}.
These bounding boxes are then passed to the CNN for prediction.
The RoIs that have been predicted as capillaries have a green bounding box around each one of them along with a black line to highlight the shape of the capillary. This is shown in Figure~\ref{example_image}e and the original image in Figure~\ref{example_image}f. 
The number of pixels within the encapsulated black contour line is summed up and divided by the total number of pixels resulting in the value of the capillary density.

Bounding boxes are formed around the predicted bounding 
boxes using the OpenCV contour method ~\cite{opencv}.
These bounding boxes are then passed to the CNN for prediction.
The RoIs predicted as capillaries have a green bounding box around them and a black line to highlight the capillary shape. This is shown in Figure~\ref{example_image}c and Figure~\ref{example_image}d. 
The number of pixels within the encapsulated black contour line is summed up and divided by the total number of pixels resulting in the value of the capillary density.

The CNN consisted of three blocks of Conv2D. 
The first Conv2D consisted of 32 filters, the second Conv2D 
consisted of 64 filters, and the third Conv2D consisted of 128 filters with a block of Maxpooling2D. 
All the Conv2D blocks have a filter of 3x3 shape.
Two dense layers of 128 neurons follow the Conv2D blocks, 64 neurons, and two neurons.
The Rectified Linear Unit (ReLu)~\cite{relu} activation function 
is used for the whole network except the last neuron layer, 
which used a softmax activation function~\cite{nwankpa2018activation}. 
This network has been trained on $\sim$11,000\ images of 
capillaries captured by trained professionals in a clinical
setting\footnote{The sponsoring company provided a device that 
was used to capture this data.}. 
The details and the specificity of the algorithms and data 
used to train the algorithm can be found in a previous paper 
by the same authors \cite{helmy2021capillarynet}.

\subsection{The Parallel System part of the Proposed System}

\begin{figure}[t]
\center
\begin{tabular}{c}
\includegraphics[scale=0.15]{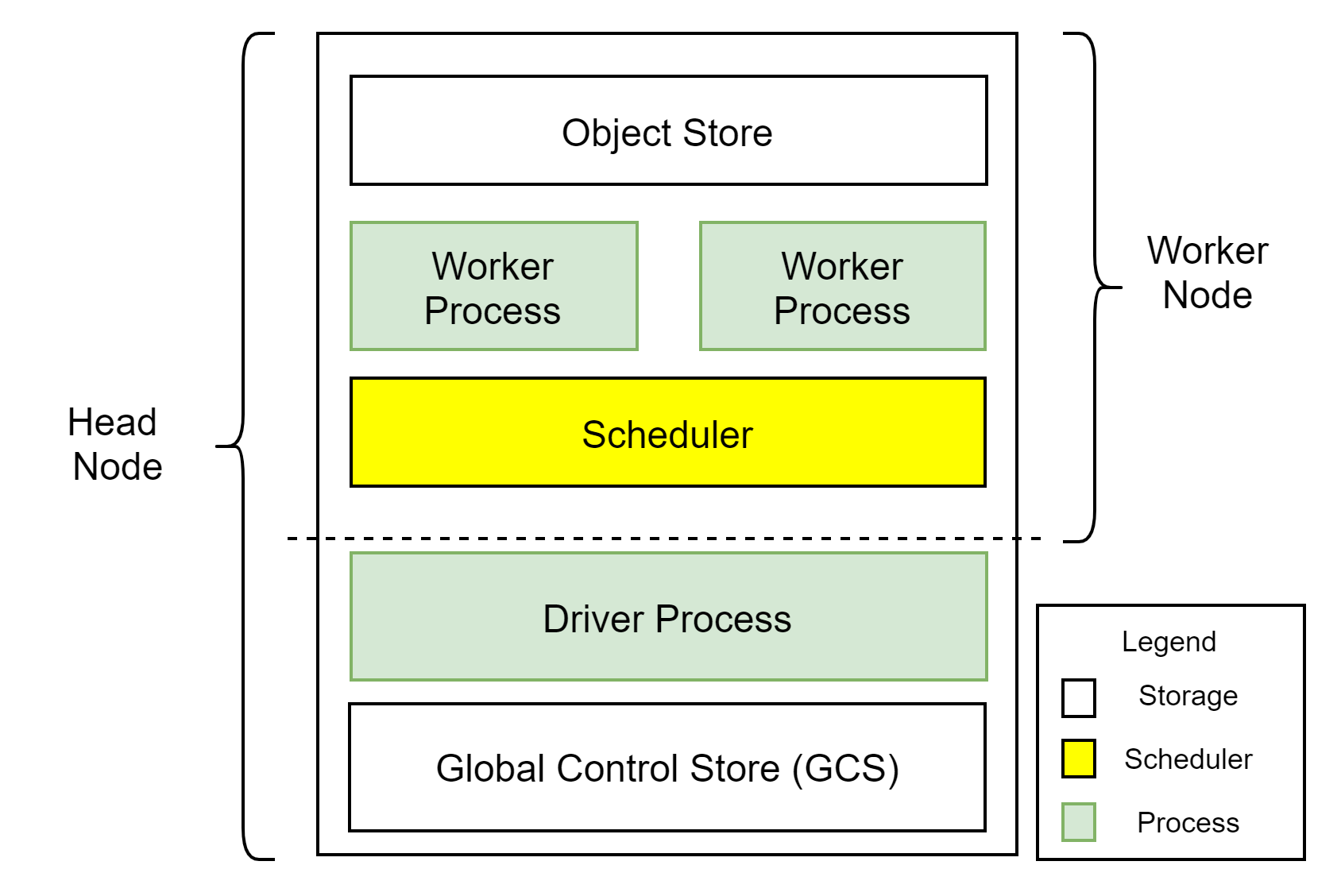}
\end{tabular}
\caption{A breakdown of the building blocks used to built \XPTO{}}
\label{workernodes} 
\end{figure}

\begin{figure}[t]
\center
\begin{tabular}{c}
\includegraphics[scale=0.2]{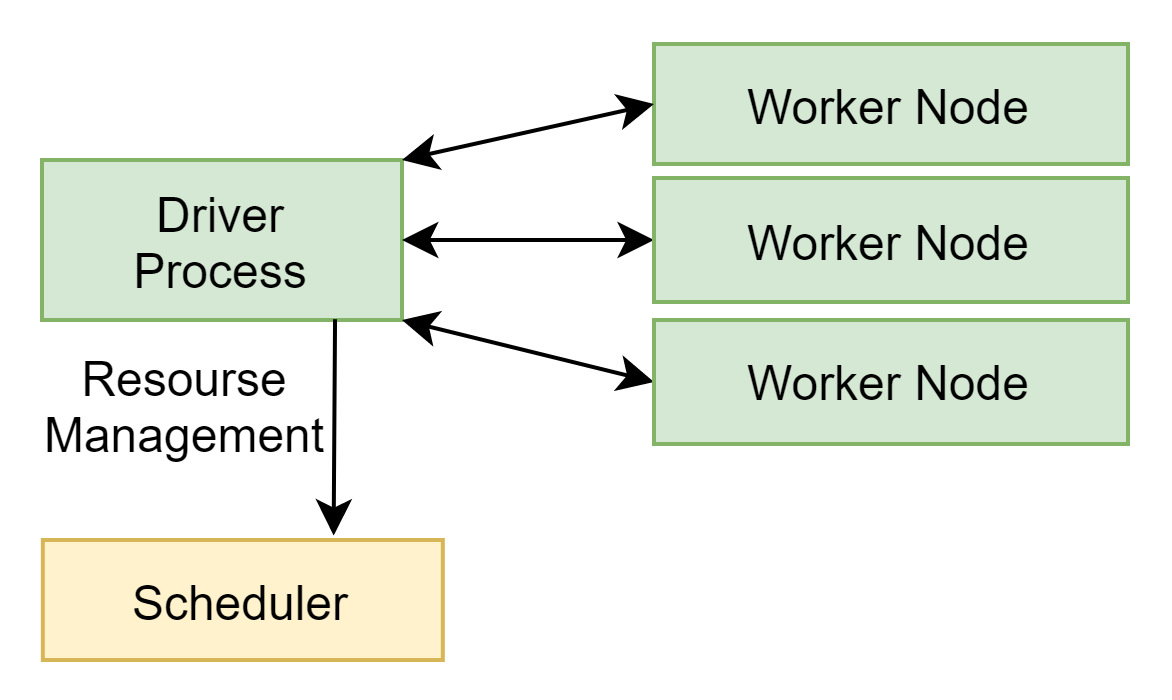}
\end{tabular}
\caption{The data flow view of how the driver process coordinates 
with the driver process and the workers}
\label{raytask} 
\end{figure}

This architecture has two types of nodes: the worker 
nodes and a head node. A worker node consists of the worker process(es), the scheduler, and the object-store.
A worker node and a head node anatomy are shown in 
Figure~\ref{workernodes} and the data flow within the components 
is shown in Figure~\ref{raytask}.

A worker process encapsulates the code to be executed and 
is responsible for task submission and execution of tasks.
In our system, the worker node encapsulates the deep learning
algorithms. It receives the image to be analyzed and replies 
whether this image contains a capillary (blood vessel) or not.
The scheduler is the resource manager of the worker node. The object store stores and transfers object larger than 100KB. 
The head node has a Global Control Store (GCS) and a driver process. 
The GCS is a key-value server that contains objects, actors, and tasks. 
The driver process submits tasks to the scheduler and keeps track of the objects created with all the nodes. 
When the code is initiated, an instance of a head node is created. 
The maximum number of worker processes within this head node is based on the number of parallel modules in the architecture instantiated and the maximum number of cores. 
Each worker performs both stages: suggesting RoIs
and detecting capillaries using the CNN loaded. 
Each worker returns a single object that contains the 
frame's density value and is stored in the object-store.
The code execution of this architecture is scheduled using 
the scheduler, and the tasks are performed over a general-purpose
Remote Procedure call to the worker processes on top of the 
Python interpreter. The scheduler then communicates the results 
via an object transfer protocol. For error handling and 
fault tolerance, the scheduler retries executing it on the 
worker processor, if a task fails due to a worker process 
ending unexpectedly.

Thus, one of the main differences between the proposed system 
and the baseline parallel system is that the former uses a 
driver process to manage the workers while the latter uses a controller and a router to manage the worker's tasks.
A baseline parallel system uses some controller 
and router to prevent the worker's potential 
overloading with tasks, which can cause it to fail.
However, these two components (controller and router) can occupy up to two cores for the management of the workers without performing any code execution. While the proposed system does not reserve any cores to manage the drivers but rather re-executes the code if a 
worker fails~\cite{robertRay}.

When the code is instantiated in our proposed system, the worker node loads the CNN as a Tensorflow model. Each worker occupies a logical 
processor, thread, or core, depending on the CPU 
architecture; we assume it is a core and instantiate a worker per core.
As the number of cores increased, 
the number of images processed in parallel increased with the number of cores.

We have shown that by combining the deep neural network part
with the parallel part, we can process several 
images at the same time, suggest RoIs and predict 
whether the bounding boxes have a capillary or not. 
Furthermore, the number of frames processed in parallel 
is determined by the maximum number of cores available or the pre-defined 
the value inserted by the user (assuming it does exceed 
the number of cores available).

\section{Implementation}
\label{implementation}

Many programming languages can implement a parallel 
processing framework. 
Python is the fastest-growing programming
language~\cite{srinath2017python,saabith2019python} and the 
preferred programming language for deep learning with 
Tensorflow~\cite{raschka2020machine,stanvcin2019overview}.
This popularity stems from its design philosophy, where it emphasizes
readability and simplicity~\cite{srinath2017python}. Moreover, 
the number of libraries, various tools, and speedily expanding the industrial community supporting Python made the language
attractive~\cite{piatetsky2019python}.

Thus, the proposed package was built on top of Python 3.7~\cite{van2007python}, OpenCV 4.5.2\cite{bradski2000opencv}, 
Scikit-learn 0.18\cite{pedregosa2011scikit}, 
Ray 1.2\cite{moritz2018Ray} and Tensorflow 2.3\cite{tensorflow1}.
The coding and evaluation were done in Pycharm Professional 2021.1 
on a Windows 10 operating system. The system can be installed, 
modified, and used by following the instructions in the readme file 
on the Github repository (www.github.com/magedhelmy1/CCGRID\_2022
\_parallel\_system\_for\_image\_analysis).

To use the system, the user can clone the package from the Github repository and import it in their Python environment.

\section{Evaluation and Discussion}
\label{evalaution}
In this section, we compare the baseline serial architecture, the baseline parallel architecture, and the proposed system with each other using the following three metrics: 
execution time, speedup, and CPU usage. We show that the 
proposed Python system is 78\% faster than the serial system and 12\% faster than the baseline parallel architecture.
These three metrics are standardized markers to quantify a system performance ~\cite{navarro2014survey}.
We use these three metrics to compare our proposed 
approach to a serial and parallel system with 
the same deep neural network.
We show that the proposed system meets the requirements 
mentioned in Section~\ref{intro} and supersedes both the baseline serial system and the baseline parallel system in execution time, speedup, and CPU usage.
The proposed system, serial counterpart, 
and parallel counterpart had the same CNN model 
and microcirculation images. 
We evaluated the three systems by taking the average 
time to calculate capillary density per image for a 
set of 100 images, which is an arbitrary number we chose to reduce the margin of error and ensure our calculations' accuracy.

\subsection{Execution Time}
\label{eval-exec}

\begin{figure}
\center
\begin{tabular}{c}
\includegraphics[scale=0.5]{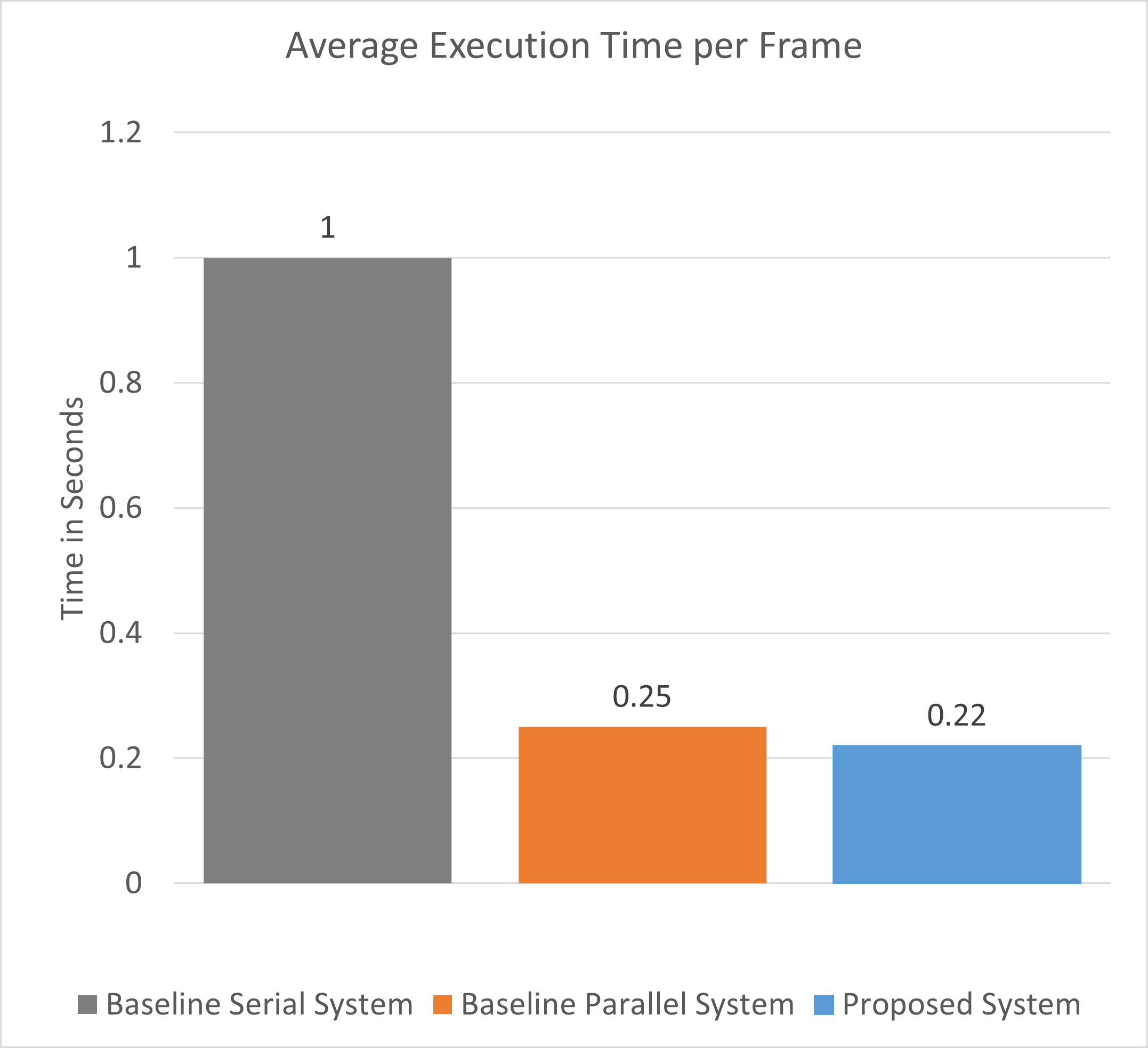}
\end{tabular}
\caption{The execution time of the proposed system against the baseline serial system and the baseline parallel system.}
\label{ET} 
\end{figure}

To calculate how much one architecture was faster compared to the other, we used Equation~\ref{eq1}, where ET denotes execution time.

\begin{equation} 
\label{eq1}
\frac{Slower ET - Faster ET}{Slower ET} = \%Faster
\end{equation}

The execution time metric measures the average time needed to 
calculate a single image's capillary density. We used 100 images in each architecture to reduce the measurement 
error margin. 
The execution time of each architecture is the following:
i) baseline serial architecture — one second per frame;
ii) baseline parallel architecture — 0.25s per frame; and 
iii) the proposed system — 0.22s per frame. 
The average values were calculated by measuring the time 
to process a frame in a set of 100 microcirculation images.
The execution time of the three architectures is presented in Figure~\ref{ET}.

In short, our results show that our proposed system 
is 12\% faster than the baseline parallel architecture and 78\% faster than its baseline serial architecture.

\subsection{Speedup}
\label{eval-speed}

\begin{figure}
\center
\begin{tabular}{c}
\includegraphics[scale=0.7]{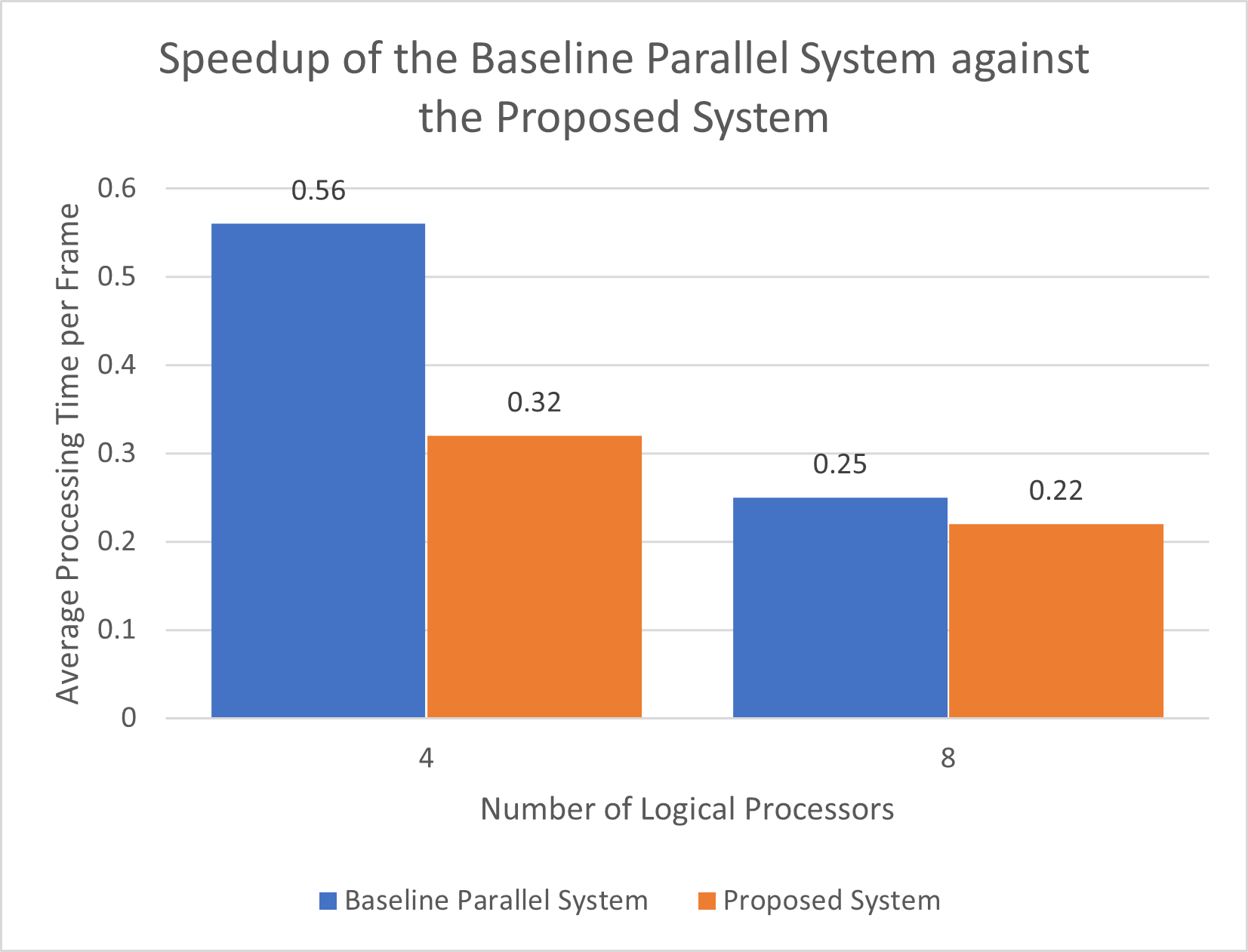} 
\end{tabular}
\caption{This graph shows the speedup of the baseline 
parallel system and \XPTO{} as the number of cores increases.}
\label{speedup} 
\end{figure}

This metric calculates the speed gain by the system as the number 
of cores increases. For the baseline serial architecture, 
the execution time is one second regardless of the number of 
cores available (indicating that the system is not 
scalable). 
The average execution time of processing one frame for the 
baseline parallel system and the proposed system is shown 
in Figure~\ref{speedup}.

The baseline parallel system 
processed a frame on average in 0.56 seconds with four cores, while the proposed system
processed a frame in 0.32 seconds. The proposed system processes 
a frame 68\% faster than the baseline serial architecture and $\sim$43\% faster than the baseline parallel architecture. 
As the number of cores doubles, the proposed system gains 
an additional $\sim$31\% during the baseline parallel 
architecture gains an additional $\sim$55\%. 
In both cases, the proposed system outperforms the baseline 
parallel architecture. 
One of the main reasons the proposed system outperformed a baseline parallel system with a master-slave architecture is that a master-slave architecture can reserve up to two cores to manage the other parallel cores. 
In contrast, the proposed system does not reserve any cores beforehand. 
In this way, we free up the computer cores to focus on processing images rather than purely handling requests. 
Thus, the proposed system gains more speedup than the baseline parallel system.
We can conclude that the proposed system has the recommended architecture 
for running deep neural networks on a single machine.

\subsection{System Resource Utilization - CPU Usage}
\label{eval-effi}

\begin{figure}
\center
\begin{tabular}{c}
\includegraphics[scale=0.4]{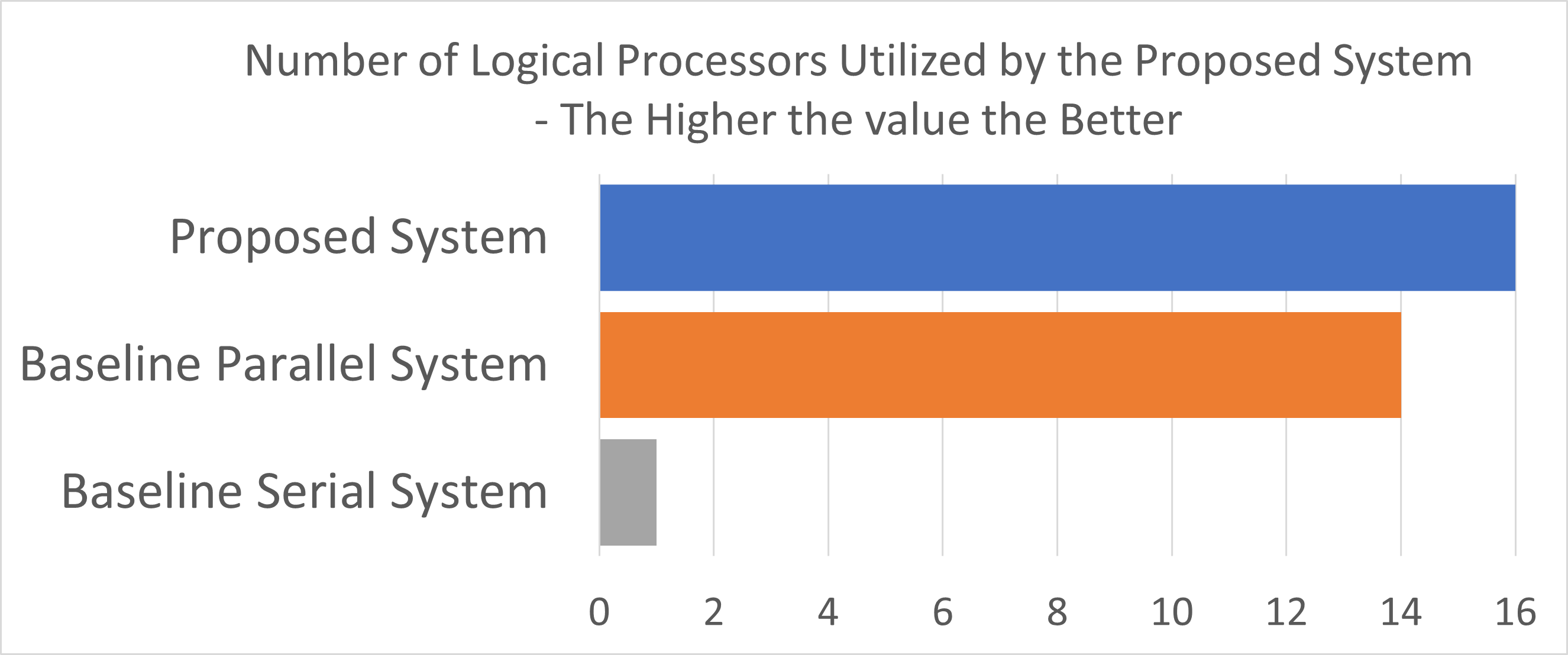}
\end{tabular}
\caption{This graph shows the number of cores used by each architecture to process a frame.
The more used at any instance the better since this shows how efficient the system is at utilizing all the resources available to it.}
\label{results_of_arch} 
\end{figure}

This metric measures the number of cores used to process the medical images using deep learning.

With the baseline serial architecture, only one cores is utilized per frame due to the Python Global Interpreter Lock's limitation. With the baseline parallel 
architecture, it is always two less than the available number of 
cores because it always reserves these two for the 
management of the parallel workers. Each cores in the proposed system is allocated a task where each task processes a frame. Thus, the proposed system is most efficient on a single machine with a multi-core. A graph showing the number of cores used by each architecture is shown in Figure~\ref{results_of_arch}.

\subsection{System Generalization}

Our system functions and classes were built using modular design patterns. 
This design philosophy means that the user can replace the DL part of our system with their algorithm by simply pointing the function in our code to their algorithm.
The details of this are highlighted in the README file in the GitHub repository. 
Thus, our package can be generalized to analyze images using a DL model of the user's 
choice in parallel. 
The system will automatically scale to the number of cores available without the user having to worry about experiencing issues with
dependency, integration, resource utilization, and speedup.

\section{Conclusion}
\label{conclusion}

This paper presented a software package that can analyze medical images using DL locally.
Our proposed system can efficiently use all local resources because it utilizes parallel execution to offset the resource-intensive demands of using a deep neural network.
The proposed system is of high clinical relevance because monitoring changes in capillary density can be used to locate early markers indicating organ failure. The severity of the change in capillary density might predict whether or not the patient survives.
Furthermore, clinical researchers do not risk uploading patient data to a third-party cloud provider to use a deep neural network to automatically analyze their images. 

Our experiments show that our system provides an optimal design for using deep learning models running on a multi-core single machine for image analysis. We benchmarked our system with a baseline serial architecture and a baseline parallel architecture using standardized evaluation metrics: execution time, speedup, and CPU usage. These metrics are used to calculate the performance of a system. Our results indicate that the proposed system is approximately 78\% faster than its baseline serial system counterpart and 12\% faster than a baseline parallel system. 

As demonstrated by our evaluation criteria, our system exhibits an acceptable industrial performance compared to the other two presented baseline systems. This argument is further strengthened because our system is currently used as a product in an industrial setting to calculate and track capillary changes in patients with pancreatitis, COVID-19, and acute heart diseases. The clinical researchers welcomed using this system to analyze their medical images locally. This acceptance was mainly due to the system reducing analysis time and removing the risks of uploading the data to a third-party cloud provider.

Our code has been made public as an open-source project in a GitHub repository for testing and usage by other clinical users. The users can import the package into their Python environment and immediately start using it. Moreover, users who can clone the code from GitHub can swap our algorithm with theirs, showing that our architecture can be generalized and utilized in the context of other use cases that require image analysis running on a CPU in near real-time. Thus, the generality of our approach can be justified by several other use cases that require image analysis.

\section*{Acknowledgment}
The authors would like to thank the Research Council of Norway for providing the necessary funds for this project. The research carried out was funded under these two projects; Industrial Ph.D. project nr: 305716 and BIA project nr: 282213.
We would also like to thank ODI Medical AS for providing the requirements, testing the system, and integrating it as part of their e-health application.

\printbibliography

\end{document}